\documentclass[10pt]{article}
\usepackage{bbm}
\usepackage[final]{graphics}
\usepackage{amsmath}
\usepackage{amsfonts,amsbsy}
\usepackage{amssymb}
\usepackage{epsfig}
\usepackage{xspace}
\usepackage{cite}

\newcommand{\npb}[3]{{\it Nucl.~Phys.~}{\bf B #1} (#2) #3}
\newcommand{\npa}[3]{{\it Nucl.~Phys.~}{\bf A #1} (#2) #3}

\newcommand{\prd}[3]{{\it Phys.~Rev.~}{\bf D #1} (#2) #3}

\def\empile#1\over#2{\mathrel{\mathop{\kern 0pt#1}\limits_{#2}}}

\def\wt#1{\widetilde{#1}}

\newcommand{\slv}{\raise.15ex\hbox{$/$}\kern-.53em\hbox{$v$}}
\newcommand{\slF}{\raise.15ex\hbox{$/$}\kern-.53em\hbox{$F$}}
\newcommand{\slL}{\raise.15ex\hbox{$/$}\kern-.53em\hbox{$L$}}
\newcommand{\slP}{\raise.15ex\hbox{$/$}\kern-.53em\hbox{$P$}}
\newcommand{\slp}{\raise.15ex\hbox{$/$}\kern-.53em\hbox{$p$}}
\newcommand{\slq}{\raise.15ex\hbox{$/$}\kern-.53em\hbox{$q$}}
\newcommand{\slR}{\raise.15ex\hbox{$/$}\kern-.53em\hbox{$R$}}
\newcommand{\slQ}{\raise.15ex\hbox{$/$}\kern-.53em\hbox{$Q$}}
\newcommand{\slK}{\raise.15ex\hbox{$/$}\kern-.53em\hbox{$K$}}
\newcommand{\slk}{\raise.15ex\hbox{$/$}\kern-.53em\hbox{$k$}}
\newcommand{\slD}{\raise.15ex\hbox{$/$}\kern-.53em\hbox{$D$}}
\newcommand{\slC}{\raise.15ex\hbox{$/$}\kern-.53em\hbox{$C$}}
\newcommand{\slA}{\raise.15ex\hbox{$/$}\kern-.53em\hbox{$A$}}
\newcommand{\slSigma}{\raise.15ex\hbox{$/$}\kern-.53em\hbox{$\Sigma$}}
\newcommand{\slpartial}{\raise.15ex\hbox{$/$}\kern-.53em\hbox{$\partial$}}
\newcommand{\slcalP}{\raise.15ex\hbox{$/$}\kern-.63em\hbox{$\cal P$}}

\def\p{{\boldsymbol p}}

\def\k{{\boldsymbol k}}

\def\x{{\boldsymbol x}}
\def\y{{\boldsymbol y}}

\def\r{{\boldsymbol r}}

\catcode`\@=11

\newcount\@tempcntc
\def\@citex[#1]#2{\if@filesw\immediate\write\@auxout{\string\citation{#2}}\fi
  \@tempcnta\z@\@tempcntb\m@ne\def\@citea{}\@cite{%
        \@for\@citeb:=#2\do%
    {\@ifundefined{b@\@citeb}%
        {\@citeo\@tempcntb\m@ne\@citea%
                \def\@citea{,\penalty\@m\ }{\bf ?}\@warning%
                {Citation `\@citeb' on page \thepage \space undefined}}%
        {\setbox\z@\hbox{\global\@tempcntc0\csname b@\@citeb\endcsname\relax}
     \ifnum\@tempcntc=\z@ \@citeo\@tempcntb\m@ne%
       \@citea\def\@citea{,\penalty\@m}%
       \hbox{\csname b@\@citeb\endcsname}%
     \else%
      \advance\@tempcntb\@ne%
      \ifnum\@tempcntb=\@tempcntc%
      \else\advance\@tempcntb\m@ne\@citeo%
      \@tempcnta\@tempcntc\@tempcntb\@tempcntc\fi\fi}}\@citeo}{#1}}%

\def\@citeo{\ifnum\@tempcnta>\@tempcntb\else\@citea
  \def\@citea{,\penalty\@m}%
  \ifnum\@tempcnta=\@tempcntb\the\@tempcnta\else
   {\advance\@tempcnta\@ne\ifnum\@tempcnta=\@tempcntb \else
\def\@citea{--}\fi
    \advance\@tempcnta\m@ne\the\@tempcnta\@citea\the\@tempcntb}\fi\fi}

\catcode`\@=12

\begin{document}

\title{\bf Limiting fragmentation in hadron-hadron\\ collisions at high energies}
\author{Fran\c cois Gelis$^{(1)}$, Anna M. Sta\'sto$^{(2,3)}$, Raju Venugopalan$^{(2)}$}
\maketitle
\begin{center}
\begin{enumerate}
\item Service de Physique Th\'eorique (URA 2306 du CNRS)\\
  CEA/DSM/Saclay, B\^at. 774\\
  91191, Gif-sur-Yvette Cedex, France
\item Brookhaven National Laboratory,\\
  Physics Department, Nuclear Theory,\\
  Upton, NY-11973, USA
\item H.~Niewodnicza\'nski Institute of Nuclear Physics,\\
  Polish Academy of Sciences, \\
  ul. Radzikowskiego 152, 31-342 Krak\'ow, Poland
\end{enumerate}
\end{center}

\maketitle

\begin{abstract}
  Limiting fragmentation in proton-proton, deuteron-nucleus and
  nucleus-nucleus collisions is analyzed in the framework of the
  Balitsky-Kovchegov equation in high energy QCD. Good agreement with
  experimental data is obtained for a wide range of
  energies. Further detailed tests of limiting fragmentation at RHIC
  and the LHC will provide insight into the evolution equations for high
  energy QCD.
\end{abstract}


\section{Introduction}
The hypothesis of limiting fragmentation in high energy hadron-hadron
collisions was suggested nearly four decades ago \cite{BCYY}. This
hypothesis states that the produced particles, in the rest frame of one of
the colliding hadrons, will approach a limiting distribution.  These
universal distributions describe the momentum distributions of the 
fragments of the other hadron. Central to the original hypothesis of the 
limiting fragmentation in Ref.~\cite{BCYY} was the assumption that the  total
hadronic cross sections would become constant at large
center-of-mass energy.  If this occurred, the excitation and break-up
of a hadron would be independent of the center-of-mass energy and
distributions in the fragmentation region would approach a limiting
curve.

We now know that the total hadronic cross-sections are not constant at
high energies. Instead, to the highest energies achieved, they grow
slowly with the center-of-mass energy $\sqrt{s}$, with favored
functional forms being either a power law behavior $\sigma(s) \propto
s^\alpha$ with $\alpha\approx 0.08$~\cite{DonnachieLandshoff} or a
$\sigma(s) \propto\ln^2(s)$ \cite{BlockHalzen}. A measurement of the
total cross-section at the LHC \cite{TOTEM} will further help
constrain the possible functional forms.

Even though the cross-sections are not constant, limiting
fragmentation appears to have a wide regime of validity. It was
confirmed experimentally in $p\bar{p},pA,\pi A$ and nucleus-nucleus
collisions at high energies \cite{SPS,FERMILAB,BRAHMS,PHOBOS}. More
recently, the BRAHMS and PHOBOS experiments at the Relativistic Heavy
Ion Collider (RHIC) at Brookhaven National Laboratory (BNL) have
performed detailed studies of the pseudo-rapidity distribution of the
produced charged particles $dN_{\rm ch}/d\eta$ for a wide range ($-5.4
< \eta < 5.4$) of pseudo-rapidities, and for several center-of-mass
energies ($\sqrt{s_{_{NN}}}=19.6, 62.4, 130 \mbox{\ and\ } 200 \, {\rm
GeV}$) in nucleus-nucleus (Au-Au and Cu-Cu) and deuteron-nucleus
(d-Au) collisions. Results for pseudo-rapidity distributions have also been obtained over a limited kinematic 
range in pseudo-rapidity by the STAR experiment at RHIC~\cite{STAR}. These measurements in A-A and d-A collisions were
performed for several centralities.  In addition to the d-Au and A-A
data, there are $pp$ data at $\sqrt{s}=200 \; {\rm GeV}$ and also at
$\sqrt{s}=410$~GeV. (In the near future, data at $\sqrt{s}=500$ GeV may
become available.)  These measurements have opened a
new and precise window on many scaling phenomena glimpsed at lower
energies. In particular, they have performed detailed studies of the
limiting fragmentation phenomenon.  The pseudo-rapidity distribution
$\frac{dN}{d\eta'}$ (where $\eta'\equiv\eta-Y_{\rm beam}$ is the
pseudo-rapidity shifted by the beam rapidity $y_{beam}=\ln
\sqrt{s}/m_p$) is observed to become independent of the center-of-mass
energy $\sqrt{s}$ in the region around $\eta'\sim 0$
\begin{equation}
\frac{dN_{\rm ch}}{d\eta'}(\eta',\sqrt{s},b)
\equiv
\frac{dN_{\rm ch}}{d\eta'}(\eta',b)\; ,
\end{equation}
where $b$ is the impact parameter.
 
It is worth noting that this scaling is in a strong disagreement with
boost invariant scenarios which predict a fixed fragmentation region
and a broad central plateau growing with energy.
It would therefore be desirable to understand the nature of hadronic
interactions that lead to limiting fragmentation and the
deviations away from it. In a recent article, Bia\l{}as and
Je\.zabek~\cite{BialasJezabek}, argued that some qualitative features of
limiting fragmentation can be explained in a two-step model involving
multiple gluon exchange between partons of the colliding hadrons and
the subsequent radiation of hadronic clusters by the interacting
hadrons.  In this paper, we compute limiting
fragmentation within the Color Glass Condensate (CGC)
approach~\cite{CGC} to high energy hadronic collisions--its relation
to the Bia\l{}as-Je\.zabek model will be addressed briefly later.

In the CGC formalism, gluon production in the limiting fragmentation
region can be described, at leading order, in the framework of
$k_\perp$-factorization. Under this assumption, the inclusive
cross-section for produced gluons is expressed as the convolution of
the ``unintegrated parton distributions" in the projectile and target
respectively~\footnote{Here and in the following, we call the
``projectile'' the nucleus $A$ which is probed at large $x_1$, and
``target'' the nucleus $B$ which is probed at small values of
$x_2$. Of course, our choice of semantics, reminiscent of 
what is used in fixed target experiments, is somewhat arbitrary in
collider experiments where the lab frame and the center of mass frame
are identical.},
\begin{equation}
\phi_{_A}(x_1,k_{1\perp}\equiv|\k_{1\perp}|)\;\phi_{_B}(x_2,k_{2\perp}\equiv|\k_{2\perp}|)\;
\delta^{(2)}(\p_\perp-\k_{1\perp}-\k_{2\perp})\; ,
\label{eq:2phis}
\end{equation}
times a (transverse momentum dependent)  vertex squared. As we shall
emphasize later, the name ``unintegrated parton distributions" though
conventional is somewhat imprecise because the objects $\phi_{_{A,B}}$
that enter in this formula differ from the expectation value of the
gluon number operator.  In eq.~(\ref{eq:2phis}), $x_1$ and $x_2$
are the longitudinal momentum fractions of the gluons probed in the
projectile and target, namely,
\begin{equation}
x_1 \equiv {p_\perp\over m_{_N}}\, e^{y-Y_{\rm beam}}\; ,\quad x_2
\equiv {p_\perp\over m_{_N}}\,e^{-y-Y_{\rm beam}} \; ,
\label{eq:kinematics}
\end{equation}
where $Y_{\rm beam} \equiv \ln(\sqrt{s}/m_{_N})$ is the beam rapidity,
$m_{_N}$ is the nucleon mass, and
$\p_\perp=\k_{1\perp}+\k_{2\perp}$ is the transverse momentum of
the produced gluon. We should emphasize that the kinematics here is
the $2\rightarrow 1$ eikonal 
kinematics, which provides the leading contribution to gluon
production in the CGC picture. 

As we will discuss further later, $k_\perp$-factorization actually
works best in the limiting fragmentation kinematics where $x_1 \geq
0.1$ and $x_2 \ll 1$ (or vice versa).  Clearly,
$k_\perp$-factorization tells us that the cross-sections depend in
general on both $y-Y_{\rm beam}$ and $y+ Y_{\rm beam}$. However,
target limiting fragmentation implies a dependence on solely $y-Y_{\rm
beam}$ of the spectrum of produced particles integrated over the
transverse momentum. As we shall see in section \ref{sec:2}, such a
scaling emerges in a straightforward manner in the
$k_\perp$-factorization framework if a) 
the typical transverse momentum in the projectile is much smaller than the typical transverse momentum
in the target, and b) if unitarity is preserved in the evolution of the
target with $x_2$. These two conditions are naturally fulfilled in the
CGC picture when $x_2$ becomes so small that the target reaches the
``black disk'' limit. More importantly, by studying how this limit is
reached, one gets some insight into systematic deviations away from the 
limiting fragmentation.

The dynamical evolution of the unintegrated distributions with $x$ are
described in the CGC by the renormalization group equations called the
JIMWLK equations~\cite{JIMWLK}. These equations, more generally,
describe the $x$ evolution of $n$-point parton correlation functions.
The dynamics in the evolution at high parton densities is
characterized by a ``saturation momentum" $Q_s(x)$. This scale is the
typical transverse momentum of partons in the hadron or nuclear
wave-function\footnote{$Q_s^2$ is also related to the density of color
charges per unit of transverse area in the hadron under
consideration.}.  Partons with $p_\perp \ll Q_s$ saturate in the
wave-function with occupation numbers of order $1/\alpha_s$. The
implications of saturated or ``black disc" distributions for limiting
fragmentation were studied previously in the CGC
approach~\footnote{Limiting fragmentation of protons in the CGC was
studied previously in Ref.~\cite{Dumitru} albeit no comparisons to
data were performed in this work.} by Jalilian-Marian~\cite{Jamal} and
in the work of Kharzeev and Levin~\cite{KL} and later by Kharzeev,
Levin and Nardi \cite{KLN}. In \cite{Jamal}, it was {\it a priori}
assumed that the target is very dense and appears black to the dilute
partons in the projectile. This leads to the simple formula
\begin{equation}
\frac{dN_{\rm ch}}{d\eta} \sim \left[x_1f^A_q(x_1,\mu^2) + x_1G^A(x_1,\mu^2) \right] \; ,
\end{equation}
for the rapidity distribution of the produced hadrons in AA
collisions, where $x_1f_q^A(x_1)$ and $x_1G^A(x_1)$ are the quark and
gluon distributions in the projectile nucleus.  With a suitable choice
of a scale $\mu$ in these distributions, this prescription gave a
reasonable description of the fragmentation region.

In \cite{KL}, the $k_\perp$-factorization formalism was employed,
together with a ``saturation" ansatz for the unintegrated gluon
distribution~\footnote{This ansatz is not the same ansatz as what one
obtains in the McLerran--Venugopalan model~\cite{KKT,BGV}.}.  Our
approach to computing the inclusive gluon production in the saturation
scenario is very similar in spirit to the one pioneered in
\cite{KL,KLN}.  Here however, the unintegrated gluon distributions at
small $x$ ($x \leq 0.01$) are computed using a mean field version of
the JIMWLK equations called the Balitsky-Kovchegov (BK)
equation~\cite{Balitsky,Kovchegov}. This approximation is strictly
valid in the large $N_c$, large mass number $A$ and high energy
limit. However, the BK equation may have a wider range of
applicability beyond these asymptotic limits.  Remarkably, it has been
shown recently that the BK equation lies in the same universality
class as the Fisher-Kolmogorov-Petrovsky-Piscounov equation in
statistical mechanics, which describe a wide range of
reaction-diffusion phenomena in nature~\cite{MunPesch}.  We will
discuss results for limiting fragmentation from solutions of the BK
equation for unintegrated distributions in both the fixed and running
coupling cases.

For $x > 0.01$, the unintegrated distribution, computed using the BK
equation, is smoothly matched on to a functional form that contains
key features expected of large $x$ parton distributions. We will also
discuss what more detailed comparisons to the data can teach us about
these distributions.

This article is organized as follows. In section 2, we discuss
inclusive gluon production in high energy hadron and nuclear
collisions. In the kinematical region of interest,  
$k_\perp$-factorization is applicable; this allows us to relate the
distribution of produced gluons to the unintegrated gluon
distributions in the hadron wave-functions.  Parton-hadron
duality~\cite{DKMT-book} is assumed to compare the distribution of
produced gluons to those of hadrons. (The effect of fragmentation functions can be significant 
at higher energies--a qualitative discussion is presented in section 4.) The evolution of the unintegrated
gluon distributions (with $x_1$ and $x_2$ respectively) in the
projectile and target wave-functions is described by solutions of the
BK equation. The fixed and running coupling forms of the BK equation
and its solution are discussed in section 3.  Results for gluon
distributions obtained from numerical solutions of the BK equation are
compared to data on limiting fragmentation in section 4.  We compare
our results to data from pp, deuteron-gold and AA collisions and discuss 
their dependence on the initial conditions, running coupling effects
and on infrared regulators. Our results are summarized and open
problems emphasized in the concluding section.


\section{Inclusive particle production\\ in high energy collisions}
\label{sec:2}
At very high energies, for $ \sqrt{s} \gg Q_s \ge p_T $, gluon
production and fragmentation is the dominant mechanism for particle
production.  When the occupation number of partons is small in one of
the colliding hadrons and large in the other (as is the case in
proton-nucleus or nucleus-nucleus collisions in the fragmentation
region of one of the nuclei), the inclusive multiplicity distribution
of produced gluons can be expressed in the $k_\perp$-factorized
form~\cite{GLR,KT,BGV},
\begin{equation}
\frac{dN_{\rm g}}{dy d^2\p_\perp}
 = 
\frac{\alpha_s S_{_{AB}}}{2\pi^4 C_{_F} (\pi R_{_A}^2)(\pi R_{_B}^2)}\,
\frac{1}{p_\perp^2}
\int \frac{d^2 \k_\perp}{(2\pi)^2}\;
  \phi_{_A}(x_1,k_\perp) \,\phi_{_B}(x_2,\big|\p_\perp-\k_\perp\big|) \; .
\label{eq:ktfact}
\end{equation}
Strictly speaking, the formula, as written here, is only valid at zero impact parameter
and assumes that the nuclei  have a uniform density in the transverse plane.
Indeed, the functions $\phi_{_{A,B}}$ are defined for the entire nucleus.
Here, $S_{_{AB}}$ denotes the transverse area of the overlap region
between the two nuclei, while $\pi R_{_{A,B}}^2$ are the total
transverse area of the nuclei, and $C_{_F} \equiv (N_c^2 -1)/ 2 N_c$ is
the Casimir in the fundamental representation.  The
longitudinal momentum fractions $x_1$ and $x_2$ were defined previously
in eq.~(\ref{eq:kinematics}).

The functions $\phi_{_A}$ and $\phi_{_B}$ are obtained from the
dipole-nucleus cross-sections for nuclei $A$ and $B$ respectively,
\begin{equation}
  \phi_{_{A,B}}(x,k_\perp)\equiv
  \frac{\pi R_{_{A,B}}^2 k_\perp^2}{4\alpha_s N_c} 
\int d^2\x_\perp\; e^{i \k_\perp \cdot \x_\perp } \; 
\left<{\rm Tr}\left(U^{\dagger}(0) U(\x_\perp)\right)\right>_{_Y} \; ,
\label{eq:unintBGV}
\end{equation}
where $Y\equiv\ln(1/x)$ and where the matrices $U$ are adjoint Wilson
lines evaluated in the classical color field created by a given
partonic configuration of the nuclei $A$ or $B$ in the infinite
momentum frame. For a nucleus moving in the $-z$ direction, they are
defined to be
\begin{eqnarray}
U(\x_\perp)\equiv T_+ \exp\left[-ig^2\int\limits_{-\infty}^{+\infty}
dz^+ \frac{1}{{\nabla}_{\perp}^2}\,\rho(z^+,\x_\perp)\cdot T\right] \; .
\label{eq:pathGlue}
\end{eqnarray}
Here the $T^a$ are the generators of the adjoint representation of
$SU(N_c)$ and $T_+$ denotes the ``time ordering'' along the
$z^+$ axis. $\rho_a(z^+,\x_\perp)$ is a certain configuration of the
density of color charges in the nucleus under consideration, and the
expectation value $\big<\cdots\big>$ corresponds to the average over
these color sources $\rho_a$. In the McLerran-Venugopalan (MV)
model~\cite{MV}, where no quantum evolution effects are included, the
$\rho$'s have a Gaussian distribution, with a 2-point correlator given
by $\left<\rho_a(0)\rho_b(\x_\perp)\right>= \mu_{_A}^2
\delta_{ab}\delta^{(2)}(\x_\perp-\y_\perp)$, where $\mu_{_A}^2 \equiv
\frac{A}{2\pi R_{_A}^2}$.  This determines $\phi_{_{A,B}}$
completely~\cite{KT,BGV}, since the 2-point correlator is all we need
to know for a Gaussian distribution.  As we will discuss later, the
small $x$ quantum evolution of the correlator in
eq.~(\ref{eq:unintBGV}) is given by the BK equation.

These distributions $\phi_{_{A,B}}$, albeit very similar to the
canonical unintegrated gluon distributions in the hadrons, should not
be confused with the latter~\cite{KKT,BGV}. However, at large
$k_\perp$ ($k_\perp \gg Q_s$), they coincide with the usual
unintegrated gluon distribution and this determines their
normalization\footnote{The unintegrated gluon distribution here is
  defined such that the proton gluon distribution, to leading order,
  satisfies
$$xG_p(x,Q^2) = {1\over 4\pi^3}\,\int_0^{Q^2} dl_\perp^2
\phi_p(x,l_\perp).$$ This normalization is different from
Ref.~\cite{KKT} - we have checked however that, when appropriately
normalized, our expression for the inclusive gluon distributions is
identical to that of Ref.~\cite{KKT}.}.

The $k_\perp$-factorized expression in eq.~(\ref{eq:ktfact}) is only
valid for inclusive gluon production when one of the hadrons is dilute
and the other is dense. In the CGC framework, this means that we keep
only terms of orders ${\cal O}(\rho_{_A}/k_\perp^2)$ and ${\cal
O}(\rho_{_B}^n/k_\perp^{2n})$ in the amplitudes, if $A$ and $B$ are
the dilute and dense hadrons respectively. This factorization is
therefore applicable for proton-proton collisions, or nucleus-nucleus
collisions in the fragmentation region of one of the projectiles.
Clearly, it applies as well to proton/deuteron--gold collisions.
$k_\perp$-factorization breaks down in kinematic regions that do not
satisfy this dilute--dense criterion.  It particular, it is not a good
approximation in the central rapidity region. Although there are some
analytical attempts to address these violations of
$k_\perp$-factorization~\cite{Balitsky2}, these have been computed
only numerically~\cite{KNV} thus far. For quark production,
$k_\perp$-factorization is broken already at leading
order~\cite{BGV2}. The magnitude of their breaking has been quantified
recently~\cite{HFR}. Here we will consider only the
$k_\perp$-factorized expression of eq.~(\ref{eq:ktfact}) with the
understanding that this expression likely has significant corrections
at central rapidities. These will be discussed further later in the
paper.

From eq.~(\ref{eq:ktfact}), it is easy to see how limiting
fragmentation emerges in the limit where $x_2\ll x_1$. In this
situation, the typical transverse momentum $k_\perp$ in the projectile
is much smaller than the typical transverse momentum
$\big|\p_\perp-\k_\perp\big|$ in the target, because these are
controlled by saturation scales evaluated respectively at $x_1$ and at
$x_2$. Therefore, at sufficiently high energies, it is legitimate to approximate
$\phi_{_B}(x_2,\big|\p_\perp-\k_\perp\big|)$ by
$\phi_{_B}(x_2,p_\perp)$. When we integrate the gluon distribution
over $\p_\perp$, we thus obtain
\begin{eqnarray}
\frac{dN_{\rm g}}{dy}
&=&
\frac{\alpha_s S_{_{AB}}}{2\pi^4 C_{_F} (\pi R_{_A}^2)(\pi R_{_B}^2)}\,
\int \frac{d^2\p_\perp}{p_\perp^2}\phi_{_B}(x_2,p_\perp)
\int \frac{d^2 \k_\perp}{(2\pi)^2}\;
\phi_{_A}(x_1,k_\perp)
\nonumber\\
&=&
\frac{S_{_{AB}}}{\pi^3 R_{_A}^2}
\int\frac{d^2 \k_\perp}{(2\pi)^2}\;
\phi_{_A}(x_1,k_\perp) \simeq \frac{S_{_{AB}}}{\pi R_{_A}^2} x_1g(x_1,\mu^2\simeq Q_s^2(x_2))\; .
\label{eq:dngdyaprox}
\end{eqnarray}
This expression, in the stated approximation, is nearly independent of
$x_2$ and therefore depends only weakly on on $y-Y_{\rm beam}$.  To
see this more clearly, note that to obtain the second line in the
above expression, we used eq.~(\ref{eq:unintBGV}) and the fact that
the Wilson line $U$ is a unitary matrix. The details of the evolution
are therefore not important to achieve limiting fragmentation, only
that the evolution equation preserve unitarity. The residual
dependence on $x_2$ comes from the the upper limit $\sim Q_s^B(x_2)$
of the integral in the second line.  This ensures the applicability of
the approximation that led to the expression in the second line
above. The integral over $\k_\perp$ gives the integrated gluon
distribution in the projectile, evaluated at a resolution scale of the
order of the saturation scale of the target. Therefore, the residual
dependence on $y+Y_{\rm beam}$ arises only via the scale dependence of
the gluon distribution of the projectile. This residual dependence on $y+Y_{\rm beam}$ is very weak at large
$x_1$ because it is the regime where Bjorken scaling is observed. 

The formula in eq.~\ref{eq:dngdyaprox} was used previously in Ref.~\cite{Jamal}. The nuclear gluon distribution here is determined by 
global fits to deeply inelastic scattering and Drell-Yan data. We note that 
the glue at large $x$ is very poorly constrained at present~\cite{NPD}. The approach of
Bialas and Jezabek \cite{BialasJezabek} also amounts to using a
similar formula, although convoluted with a fragmentation function
(see eqs.~(1), (4) and (5) of \cite{BialasJezabek} -- in addition,
both the parton distribution and the fragmentation function are
assumed to be scale independent in this approach). We will discuss the effect of fragmentation functions later in 
our discussion. 

Though limiting fragmentation can be simply understood as a
consequence of unitarity in the high energy limit, what may be more
compelling are observed deviations from limiting fragmentation and how
these vary with energy. These deviations may probe more deeply our
understanding of the dynamics of both large $x$ and small $x$ modes in hadronic wavefunctions. In
particular, in the small $x$ case, it may provide further insight into the
renormalization group equations that, while trivially preserving
unitarity, demonstrate interesting pre-asymptotic behavior. These
concerns provide the motivation for this detailed study with the
Balitsky-Kovchegov renormalization group equation--to be discussed in
the following section.


\section{Balitsky-Kovchegov equation}
\label{sec:BK}
We begin by briefly recapitulating key features of the
Balitsky-Kovchegov (BK) equation and its
solution~\cite{Balitsky,Kovchegov}. Readers are referred to recent
review literature on the subject for a more detailed
discussion~\cite{CGC,JamalYuri}.  The BK equation is a non-linear
evolution equation in rapidity $Y=\ln(1/x)$ for the forward scattering
amplitude of a quark-antiquark dipole scattering off a target in the
limit of very high center-of-mass energy $\sqrt{s}$. It was originally
derived, within the dipole picture (which assumes the large $N_c$
limit) at small values of Bjorken $x$, by taking into account multiple
rescatterings of the $q\bar{q}$ dipoles off a dense nuclear
target. The BK equation for the amplitude is equivalent to the
corresponding JIMWLK equation~\cite{JIMWLK} of the Color Glass
Condensate, in a mean field (large $N_c$ and large $A$) approximation
where higher order dipole correlators are neglected. The
parametrically suppressed $N_c$ and $A$ contributions can, in
principle, be computed by solving the JIMWLK equation. In momentum
space, the BK equation takes the form~\footnote{Here we present the
form of the equation for the case where the forward scattering
amplitude is independent of the impact parameter. A numerical study of
the impact parameter dependence of the BK equation has been performed
previously~\cite{GBMS}; it will be considered in future as an
extension to this work.}
\begin{equation}
\frac{\partial {\widetilde{T}}(k_\perp,Y)}{\partial Y}
=
\overline{\alpha}_s (K\otimes \widetilde{T})(k_\perp,Y)
-
\overline{\alpha}_s \widetilde{T}^2(k_\perp,Y)\; ,
\label{eq:kov}
\end{equation}
where we denote $\overline{\alpha}_s\equiv \alpha_s N_c/\pi$.  The
operator $K$ is the well known BFKL kernel in momentum
space~\cite{BFKL}, whose action on the function $\widetilde{T}$ is
given by
\begin{equation}
(K\otimes \widetilde{T})(k_\perp,Y)
\equiv
\int\limits_0^{+\infty}
\frac{d(k^{\prime 2}_\perp)}{k^{\prime 2}_\perp}\;
\left\{
\frac{k^{\prime 2}_\perp\widetilde{T}(k^\prime_\perp,Y)
-k_\perp^2\widetilde{T}(k_\perp,Y)}{\big|k_\perp^2-k^{\prime 2}_\perp\big|}
+
\frac{k_\perp^2\widetilde{T}(k_\perp,Y)}{\sqrt{4k_\perp^{\prime 4}+k_\perp^4}}
\right\}\; .
\end{equation}
  The function $\widetilde{T}(k_\perp,Y)$ is the Bessel-Fourier
transform of the dipole-target scattering amplitude $T(r_\perp,Y)$:
\begin{equation}
\widetilde{T}(k_\perp,Y)
=
\int\limits_0^{+\infty} \frac{dr_\perp}{r_\perp}  \, 
J_0(k_\perp r_\perp) \,  T(r_\perp,Y)\; ,
\label{eq:unint-gluon}
\end{equation}
where $r_\perp$ is the size of the $q\bar{q}$ dipole and $k_\perp$ is
its conjugate transverse momentum.  The dipole amplitude $T$ is
defined in terms of the correlator of two Wilson lines of gauge fields
in the target as 
 \begin{equation}
 T(r_\perp,Y)
=
1- \frac{1}{N_c}{\rm Tr} \left<
 {\tilde U}^\dagger(0) {\tilde U}(\r_\perp)\right>_{_Y} \; ,
 \label{eq:dipole}
 \end{equation}
where we have assumed translation invariance in the transverse plane in
order to set the quark transverse coordinate to $0$.  Here ${\tilde
U}$ is a Wilson line in the fundamental representation, obtained by
replacing in eq.~(\ref{eq:pathGlue}) the adjoint generators $T^a$ by
the generators $t^a$ in the fundamental representation.
 
However, to evaluate the ``unintegrated gluon distribution'' defined
in eq.~(\ref{eq:unintBGV}), one needs to know the rapidity dependence
of the correlator of two Wilson lines in the {\sl adjoint
representation}, $\left<U(0)U^\dagger(\r_\perp)\right>_{_Y}$. This can
be obtained as follows. In the CGC framework, the weight functional
for the color sources in generating functional can be expressed in the
large $N_c$ and large $A$ limit as a non-local Gaussian distribution
of color sources. In this limit, one can obtain closed form
expressions for expectation values of both the fundamental and adjoint
correlators of Wilson lines~\footnote{For a detailed discussion and
relevant references, we refer the reader to Appendix A of
Ref.~\cite{BGV2}.}. One therefore has
\begin{eqnarray}
&&
{\rm Tr}\left<{\wt U}(0){\wt U}^\dagger(\r_\perp)\right>_{_Y}
=N_c \; e^{-C_{_F}\Gamma(r_\perp,Y)}\; ,
\nonumber\\
&&
{\rm Tr}\left<U(0)U^\dagger(\r_\perp)\right>_{_Y}
= N_c^2 \; e^{-C_{_A}\Gamma(r_\perp,Y)}\; ,
\label{eq:largeN-corr}
\end{eqnarray}
where $C_{_A}\equiv N_c$ is the Casimir in the adjoint
representation. The function $\Gamma$ is proportional to the variance
of the non-local Gaussian weight functional in the generating
functional and is therefore the same in both the fundamental and
adjoint cases. As $C_{_A}/C_{_F}=2$ in the large $N_c$ limit,
eqs.~(\ref{eq:largeN-corr}) and ~(\ref{eq:dipole}) give
\begin{equation}
\frac{1}{N_c}
\;
{\rm Tr}\left<U(0)U^\dagger(\r_\perp)\right>_{_Y}
=N_c \left[1-T(r_\perp,Y)\right]^2\; .
\end{equation}
Substituting the LHS side here by the RHS into
eq.~(\ref{eq:unintBGV}), we obtain
\begin{equation}
\phi_{_{A,B}}(x,k_\perp) 
= 
\frac{\pi^2 R_{_A}^2 N_c k_\perp^2}{2\,\alpha_s}
\int\limits_0^{+\infty} r_\perp dr_\perp 
\;
J_0(k_\perp r_\perp)
\,
\left[1-T_{_{A,B}}(r_\perp,\ln(1/x))\right]^2
 \; .
 \label{eq:phiadjoint}
\end{equation}

\begin{figure}[hbt]
\centerline{\epsfig{file=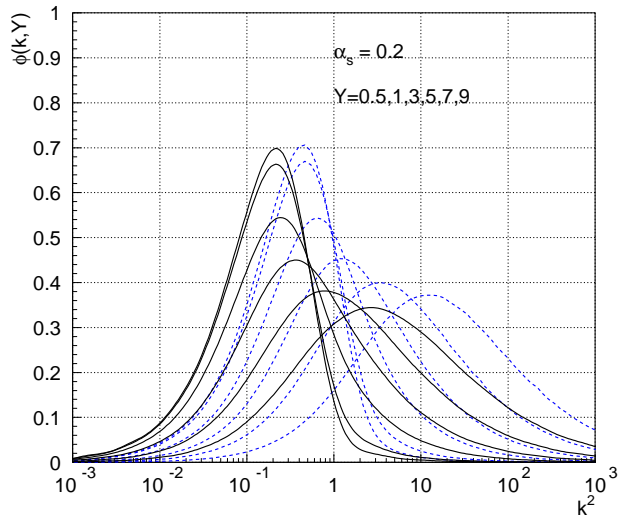,width=0.7\textwidth}}
\caption{The unintegrated gluon distribution obtained from 
a) correlators in the adjoint representation (eq.\ref{eq:phiadjoint})
(dashed lines) and b) correlators in the fundamental representation 
(eq.~\ref{eq:phifundamental}) (solid lines).}
\label{fig:1a}
\end{figure}

We digress here to note that if instead we had used the correlator of two Wilson lines in the
fundamental representation in eq.~(\ref{eq:unintBGV}), we would have
obtained the following expression for the unintegrated gluon density
\begin{equation}
\widetilde{\phi}_{_{A,B}}(x,k_\perp) 
= 
\frac{\pi^2 R_{_A}^2  k_\perp^2}{2\,\alpha_s}
\int\limits_0^{+\infty} r_\perp dr_\perp 
\;
J_0(k_\perp r_\perp)
\,
\left[1-T_{_{A,B}}(r_\perp,\ln(1/x))\right]
 \; .
 \label{eq:phifundamental}
\end{equation}
The change from the correlator of Wilson lines in the fundamental
representation to the adjoint one corresponds to the emission of the
gluon from the triple Pomeron vertex itself. This emission is not  prohibited by
the Abramovsky--Gribov--Kancheli (AGK) cutting rules~\cite{AGK} for inclusive gluon production in high energy QCD. 
It was argued previously~\cite{Braun1,Braun2} that the
numerical difference in the resulting rapidity distributions is rather
small. In figure~\ref{fig:1a}, we compare numerical solutions (to be discussed further shortly) 
for the unintegrated gluon
distribution using the correlator of Wilson lines in the
adjoint representation (eq.~\ref{eq:phiadjoint})) with the one using
the correlator of Wilson lines in the fundamental representation 
(eq.\ref{eq:phifundamental}). The shape of the distribution is very
similar, whereas the position of the peak is different. At $Y=0$ the
position of the peaks differ by the ratio of the $C_A/C_F \simeq 2$.
At higher values of $Y$ this difference increases due to the faster
evolution of the correlator in the adjoint representation.
 
The BK equation is the simplest evolution equation capturing both the
leading $\ln(1/x)$ BFKL dynamics at moderately small values of $x$, as
well as the recombination physics of high parton densities at very
small values of $x$. In the limit where the dipole scattering
amplitude is small, $T\ll 1$, the non-linear term in
eq.~(\ref{eq:kov}) can be ignored and the BK equation reduces to the
BFKL equation. In this limit, the amplitude has the solution,
\begin{eqnarray}
T(r_\perp,Y) &\approx& (r_\perp^2 Q_0^2)^{1/2}\,e^{\omega
{\overline{\alpha}_s}Y}\exp\left(-{\ln^2(1/r_\perp^2 Q_0^2)\over
2\beta {\overline{\alpha}_s} Y}\right) \nonumber \\ &=& \exp\left(
{\rho\over 2} + \omega {\overline{\alpha}}_s Y -{\rho^2\over 2\beta
{\overline{\alpha}}_s Y} \right) \; ,
\label{eq:omega}
\end{eqnarray}
where $\omega = 4 \ln 2\approx 2.77$, $\beta=28\zeta(3) \approx 33.67$
and $\rho \equiv \ln(r_\perp^2 Q_0)^2$.  If we define the saturation
condition as
\begin{equation}
T(r_\perp,Y)=\frac{1}{2}\quad\mbox{at\ \ } 
r_\perp={2\over Q_s}\; ,
\label{eq:31}
\end{equation}
one gets~\cite{MuelTrian}
\begin{equation}
Q_s^2 = Q_0^2 \,\, e^{c {\overline{\alpha}}_s Y} 
\quad\mbox{where} \; c\approx4.84 \; .
\label{eq:fixed-sat-scale}
\end{equation}
A more careful analysis of the BK-equation close to the saturation
boundary gives solution $c\approx 4.88$. In the strong field (saturation)
limit~\cite{StrongField} where $T\approx 1$, one finds
\begin{equation}
 T(r_\perp,Y) 
= 1-{\kappa}\;
\exp\left(-{1\over 4c}\,\ln^2(r_\perp^2 Q_s^2)\right) \, ,
\label{eq:34}
\end{equation}
with $c\approx 4.84$ and where $\kappa$ is an undetermined
constant. In the large $Y$ asymptotic regime, leading and
next-to-leading corrections to the form of the BK amplitude and the
saturation scale have been computed~\cite{MunPesch,Dionysis}.

The coupling constant $\overline{\alpha}_s$ in eq.~(\ref{eq:kov}) is
fixed in the leading logarithmic approximation in $ 1/x$. This leads
to a rather strong dependence of the saturation scale in the leading
logarithmic approximation: $Q_s\sim x^{-\lambda_s}$ with $\lambda_s
\approx 4.88\,\bar{\alpha}_s $. For typical values
of $\alpha_s$, of order $0.2$-$0.3$, $\lambda_s$ is at least a factor
of three higher than fits to the HERA and RHIC data would
suggest~\cite{GBW}. It is therefore desirable to include at least part
of the next-to-leading corrections in $\ln(1/x)$ to the BK
equation. Such corrections are known to reduce significantly the
Pomeron intercept in the case of the BFKL equation. 

The BK equation has been solved numerically for both fixed and running
coupling~\cite{ArmestoBraun,LevLub,GBMS,WeiRum}. These solutions have the
following features:
\begin{itemize}
\item The dipole amplitude is shown to unitarize, and the solution
  exhibits geometrical scaling.
\item The saturation scale has the behavior in
  eq.~(\ref{eq:fixed-sat-scale}) or $Q_s\sim\exp(\sqrt{\lambda \ln 1/x})$ in the fixed and running coupling cases respectively.
\item The infrared diffusion problem of the BFKL solution is cured by
  the non-linear term in the BK equation.
\item Ultraviolet diffusion is still present in the BK
  equation. However, it is attenuated by including running coupling
  effects~\cite{ArmestoBraun,WeiRum,Albacete2}.
\end{itemize}

In this work, we solved the BK equation numerically, in both fixed and running 
coupling cases, to investigate limiting fragmentation in
hadronic collisions. In the fixed coupling case, since realistic
values of $\alpha_s$ give too high a value of $\lambda_s$, we solved
the BK equation with very small values of $\alpha_s$ chosen to give
$\lambda_s$ values that are compatible with data from HERA, RHIC and
hadron colliders. (We will see that these lower values are indeed
favored by the data.) This is not unreasonable because it has been
shown~\cite{Dionysis} that resummed next-to-leading order corrections
to the BFKL equation in the presence of a unitary boundary ({\it \`a
la} BK) give values of the saturation scale $\lambda_s$ that are
nearly independent of the energy, as in the LO case, but with a much
smaller value of $\lambda_s$. 

The limited running coupling studies we performed  
did not give nearly as good agreement with the data than the fixed coupling studies with the 
lower values of $\alpha_s$. This is likely due to the fact that the energy dependence in this case 
is much too fast compared to the data. Further, since our results are sensitive to small transverse momenta, 
they will also depend strongly on how one regulates the running of $\alpha_S$ in the infrared. This requires a more 
detailed study than reported here.  We shall therefore restrict ourselves to the fixed coupling case in discussing our 
results in the following section. 

As explained before, the ``unintegrated gluon distributions''
$\phi_{_{A,B}}$ are obtained from solutions to eq.~(\ref{eq:kov}). At
small values of $x$, where the gluon density is high,
eq.~(\ref{eq:kov}) captures the essential physics of saturation; we
therefore use it to determine $\phi(x,k_\perp)$ for values of $x<x_0$
with $x_0=0.01$.  For larger values of $x$ ($x \geq 0.01$), we used
the phenomenological extrapolation\footnote{A similar extrapolation
was also used in Ref.~\cite{HFR} to study quark pair production.},
\begin{equation}
\phi(x,k_\perp) 
=  
\bigg(\frac{1-x}{1-x_0}\bigg)^{\beta} \, 
\bigg(\frac{x_0}{x}\bigg)^{\lambda_0} \, 
\phi(x_0,k_\perp)\; ,\quad x>x_0 \; .
\label{eq:extrapolx}
\end{equation}
The parameter $\beta=4$ is fixed by QCD counting
rules~\cite{counting-rules}.  We checked that the results are not
sensitive to the variation of this parameter in the range $4-5$. The
parameter $\lambda_0$ was varied between $0$ and $0.15$ in the fits to
data discussed in the following section.

The initial condition for the BK evolution, that gives
$\phi(x_0,k_\perp)$, is given by the McLerran-Venugopalan (MV)
model~\cite{MV} with a fixed initial value of the saturation scale
$Q_s^A(x_0)$. For a gold nucleus, extrapolations from HERA and
estimates from fits to RHIC data suggest that $(Q_s^A(x_0))^2\approx 2
\, {\rm GeV}^2$. The saturation scale in the proton is taken to be
$Q_s^2(x_0)= (Q_s^A(x_0))^2 \,(197/A)^{1/3}$.  For comparison, we also
considered initial conditions from the Golec-Biernat and Wusthoff
(GBW) model~\cite{GBW}. The values of $Q_s^A$ were varied in this
study to obtain best fits to the data.

\section{Results for Limiting Fragmentation}

Experimental data are presented in terms of the measured distributions
of produced charged hadrons.  Our expression in eq.~(\ref{eq:ktfact})
is for produced gluons. Rapidity distributions are dominated by low
$p_\perp$ ($p_\perp \leq Q_s$) particles; the detailed mechanism of
how such soft gluons fragment to form hadrons is unknown. However, in
$e^+ e^-$ collisions, several studies have been performed of
hadronization; it is observed that at scales comparable to $p_\perp
\sim Q_s$, the distribution of produced hadrons mirrors that of the
produced gluons. This hypothesis is known as ``parton-hadron"
duality~\cite{DKMT-book} and we shall adopt it here.\footnote{This
assumption was also made in ref.~\cite{KL} and is implicit in several
other works. Note that such an assumption may be invalidated by final
state interactions such at medium induced parton splitting, namely, ``jet
quenching''. As we shall discuss later in detail, we observe that the $p_\perp$ spectral shapes of softer partons,  
albeit one would imagine them to interact more strongly, are closer to that of observed hadrons-than those with $p_\perp > Q_s$. 
This is fortuitous for the discussion of limiting fragmentation here since $dN/dy$ is dominated by soft momenta.} We will discuss the 
effect of fragmentation functions shortly. 

Limiting fragmentation is often experimentally studied in terms of
the measured pseudo-rapidity $\eta$ of produced particles. For
massless particles, $\eta$ and $y$ are the same, but they differ for
massive particles. One obtains
\begin{equation}
y(\eta,p_\perp,m) \; = \; \frac{1}{2} \ln \left[\frac{\sqrt{m^2 +p_\perp^2 \, {\rm ch}^2 \eta}+p_\perp \, {\rm sh} \eta}{\sqrt {m^2 +p_\perp^2 \, {\rm ch}^2 \eta}-p_\perp \, {\rm sh} \eta}\right] \; .
\label{eq:yeta}
\end{equation}
We choose $m$ to be of order of $m \simeq 150 \,{\rm MeV}-300 \, {\rm
MeV}$. Our expression for $dN/dy$ of gluons has a logarithmic
divergence at low $p_\perp$; in addition to parton-hadron duality, for
self consistency, we should regulate the corresponding expression for
charged hadrons with the same mass as the one used in the conversion
from $y$ to $\eta$. We will later discuss the sensitivity of our
results to $m$.

We now have all the ingredients necessary to calculate the
 multiplicity distributions in the Color Glass Condensate
 framework. As discussed previously, experimental data on
 pseudo-rapidity distributions in the projectile fragmentation region
 scale with $\eta-Y_{\rm beam}$ for different energies.  In
 eq.~(\ref{eq:ktfact}), $\phi_{_A}(x_1)$ depends on the difference
 $y-Y_{\rm beam} $ and $\phi_{_B}(x_2)$ on the sum $y+Y_{\rm
 beam}$. To provide a sense of the $x$'s involved, consider gold-gold
 collisions at RHIC.  For $y =Y_{\rm beam}$, which lies in the
 fragmentation region, and $p_\perp \sim m_N \sim 1$ GeV , $x_1
 \approx 1$ and $x_2\approx 2.5\cdot 10^{-5}$. One is therefore
 probing very small values of $x_2$ in nuclei in this
 region\footnote{Note that $x_1$ can be larger than unity in a
 nucleus and could in principle take values up to $A$, the number of
 nucleons.}, where the
 saturation scale $Q_s(x_2)$ is rather large. As we discussed
 previously, in this kinematic region, because of unitarity, the gluon
 distribution depends only weakly on $x_2\sim \exp(-(y+Y_{\rm
 beam}))$. This qualitatively explains the scaling in the limiting
 fragmentation region.

We now turn to our results. To reiterate, they are obtained by a) solving
the BK equation to obtain eq.~(\ref{eq:unint-gluon}), and thereby eq.~(\ref{eq:phiadjoint}) for the unintegrated
distribution for $x \leq 10^{-2}$, b) using eq.~(\ref{eq:extrapolx}) to
determine the large $x$ ($x \geq 10^{-2}$) behavior, and c) substituting
these in eq.~(\ref{eq:ktfact}) to determine rapidity distribution of
gluons. The pseudo-rapidity distributions are determined by the
transformation in eq.~(\ref{eq:yeta}).

\begin{figure}[hbt]
\centerline{\epsfig{file=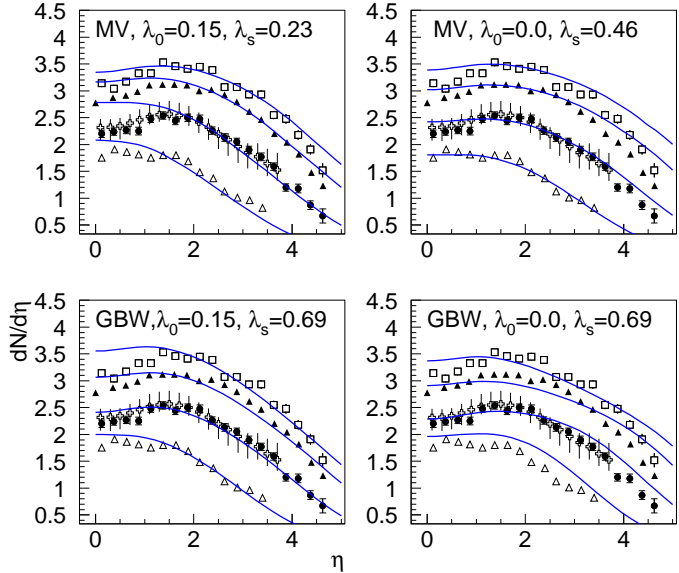,width=0.8\textwidth}}
\caption{ Pseudorapidity distributions $dN/d\eta$ for charged
particles from nucleon-nucleon collisions at UA5 energies~\cite{SPS}
$\sqrt{s}=53, 200, 546, 900 \, {\rm GeV}$ and PHOBOS data~\cite{PHOBOS} at
$\sqrt{s}=200 \, {\rm GeV}$. Upper plots: initial distribution from
the MV model, lower plots: initial distribution from the GBW
model. Left panels: $\lambda_0=0.15$, right panels $\lambda_0=0.0$.  }
\label{fig:2}
\end{figure}

In figure~\ref{fig:2} we show pseudo-rapidity distributions of the
charged particles produced in nucleon-nucleon collisions for center of
mass energies ranging from $53 \, {\rm GeV}$ to $900 \, {\rm GeV}$.
We have performed calculations for two different input distributions
at starting value of $x_0=0.01$, namely, the GBW and MV models. The
normalization has been treated as a free parameter.  Extrapolations to
large values of $x$ are performed using the formula in
eq.~(\ref{eq:extrapolx}).  Plots on the left hand side of
figure~\ref{fig:2} are obtained for $\lambda_0=0.15$ whereas the right
hand side plots are done for $\lambda_0=0.0$. The different values of
$\lambda_s$ are obtained by varying $\alpha_s$ when solving the BK
equation. 

Note that while $\lambda_s$ controls the growth of $Q_s$ with energy,
the amplitude in eq.~(\ref{eq:omega}) has the growth rate
\begin{equation}
\lambda_{\rm BK} = \frac{2.77}{4.88}\, \lambda_s \approx 0.57\,\lambda_s \, ,\nonumber
\end{equation} 
which is significantly lower. So $\lambda_s =0.46$, which gives
reasonable fits (more on this in the next paragraph) to the pp data
for the MV initial conditions, corresponds to $\lambda_{\rm BK} =
0.28$, which is close to the value for the energy dependence of the
amplitude in next-to-leading order resummed BK
computations~\cite{Dionysis} and in empirical dipole model comparisons
to the HERA data~\cite{GBW}.

We find that our computations are extremely sensitive to the
extrapolation prescription to large $x$. This  is not a surprise since we are
probing the wave-function of the projectile at fairly large values of
$x_1$. From our analysis, we see that the data naively favor a non-zero value
for $\lambda_0$ in eq.~(\ref{eq:extrapolx}).  The zero value of
$\lambda_0$ results in the distributions which, in both the MV and GBW cases give reasonable fits (albeit with 
different normalizations) at 
lower energies but systematically become harder relative to the data as the energy is increased. 
To fit the data in the MV model up to the highest UA5 energies, a lower value of $\lambda_s$ than that in the
GBW model is required. This is connected with the fact that MV model has tails
which extend to larger values in $k_\perp$ than in the GBW model. As the energy is increased, the 
typical $k_\perp\sim Q_s(x_2)$ does as well. We will return to this point shortly.

\begin{figure}[hbt]
\centerline{\epsfig{file=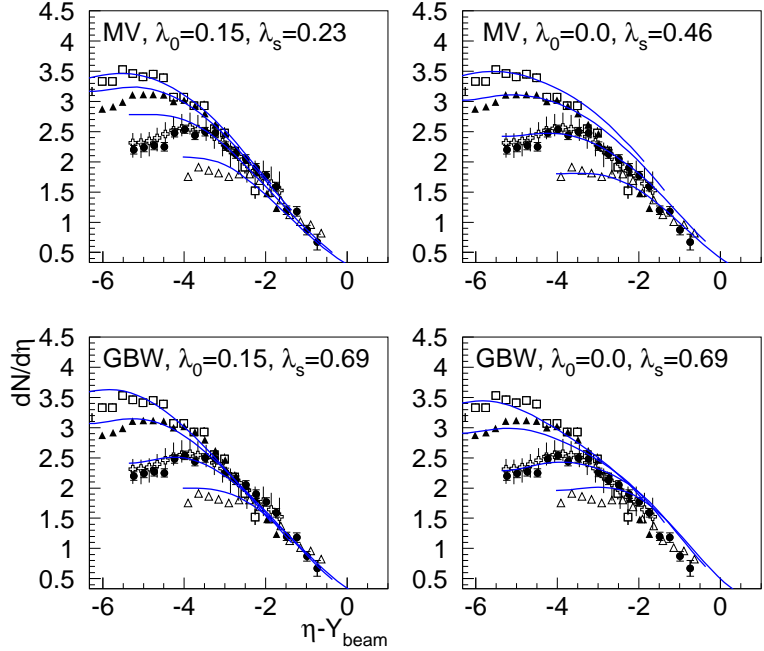,width=0.9\textwidth}}
\caption{
The same as figure~\ref{fig:2} but plotted versus $\eta'=\eta-Y_{\rm beam}$ to illustrate the region of limiting fragmentation.
}
\label{fig:3}
\end{figure}
  
In figure~\ref{fig:3} the same distributions are shown as a function
of the $\eta'=\eta-Y_{\rm beam}$.  The calculations for
$\lambda_0=0.15$, are consistent with scaling in the limiting
fragmentation region.  There is a slight discrepancy between the
calculations and the data in the mid-rapidity region. This discrepancy
may be a hint that one is seeing violations of $k_\perp$ formalism in
that regime because $k_\perp$ formalism becomes less reliable the
further one is from the dilute-dense kinematics of the fragmentation
regions~\cite{KNV,Balitsky2}. This discrepancy should grow with increasing energy. However, our parameters are not sufficiently
constrained that a conclusive statement can be made.

\begin{figure}[hbt]
\centerline{\epsfig{file=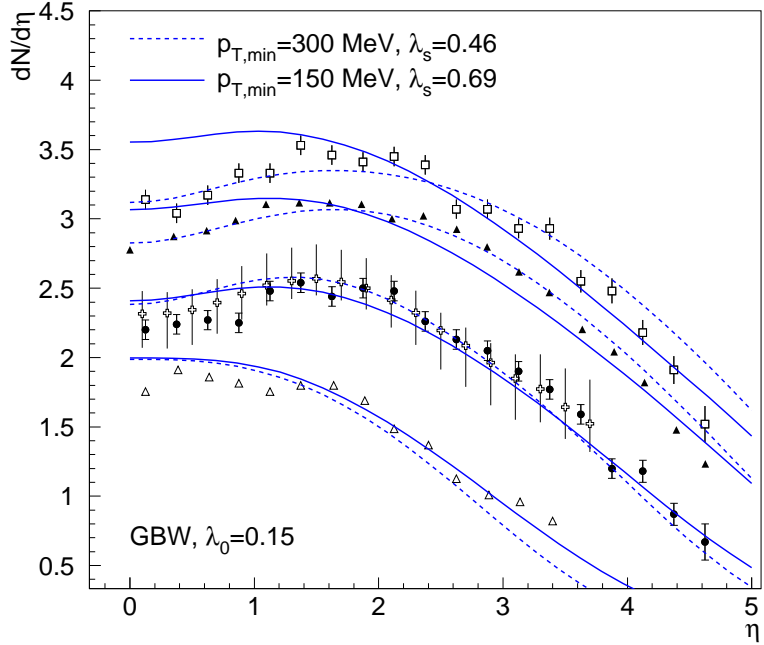,width=0.9\textwidth}}
\caption{
Calculations for different values of the infrared $p_T$ cutoff compared to the pp data in figure~\ref{fig:2}.
}
\label{fig:3a}
\end{figure}

In changing from rapidity to pseudo-rapidity distributions, one has to
use a Jacobian with a mass parameter. We have chosen this parameter,
for consistency, to be the same as our $p_\perp$ cut-off for the
integration over the transverse momentum in the formula for the gluon
production.  We have checked the sensitivity of our calculations to
variations in the $p_\perp$ cut-off $m$ in the range $150-300 \, {\rm
MeV}$. In figure~\ref{fig:3a} we show the results for the pp
collisions with two different values of $p_{\perp,{\rm min.}} \; = \; 150, 300 \,
{\rm GeV}$. The `dip' in the pseudo-rapidity distribution becomes more
or less pronounced when the parameter $m$ is decreased or increased
respectively. We note that in order to obtain a reasonable description
of the data we also had to adjust the other parameters (normalization
and $\lambda_s$).

\begin{figure}[hbt]
\centerline{\epsfig{file=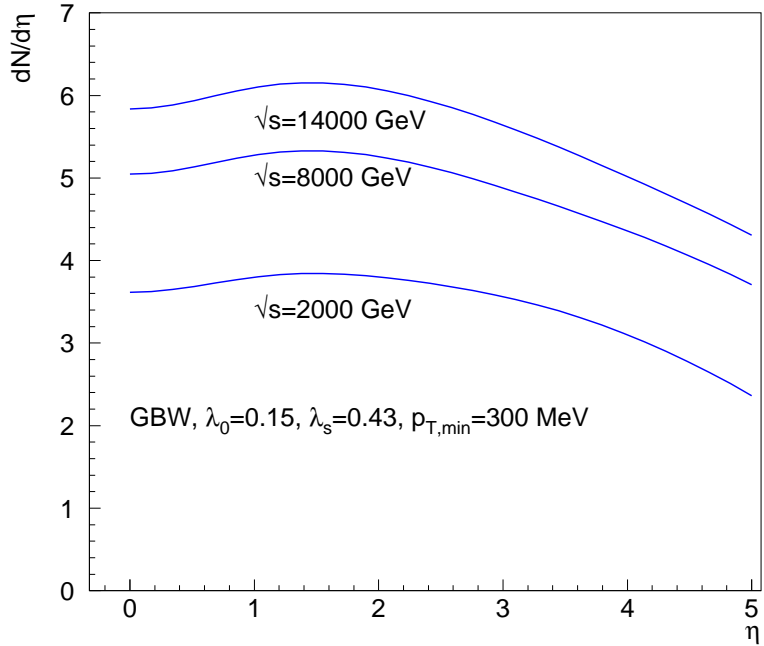,width=0.9\textwidth}}
\caption{
Predictions for higher center-of-mass energies: $\sqrt{s}=2,8,14 \, {\rm TeV}$
for GBW input model. Parameter in the large $x$ extrapolation was set to $\lambda_0=0.15$.
}
\label{fig:4}
\end{figure}

In figure~\ref{fig:4} we show the extrapolation to higher energies, in
particular the LHC range of energies for the calculation with the GBW
input.  We observed previously that the MV initial distribution, when
evolved to these higher energies, gives a rapidity distribution which
is very flat in the $-5<\eta<5$ regime. We noted that this is because
the average transverse momentum grows with the energy giving a
significant contribution from the high $k_\perp$ tail of the
distribution in the MV input in eq.~(\ref{eq:extrapolx}).  

The effect of fragmentation functions on softening the spectra in the limiting fragmentation region can 
be simply understood by the following qualitative argument. The inclusive hadron distribution can be expressed as 
\begin{equation}
\frac{dN_h}{d^2 p_\perp dy} = \int_{z_{\rm min.}}^1 \frac{dz}{z}\, \frac{dN_g}{d^2 q_\perp dy}\, D_{g\rightarrow h} \left(z=\frac{p_\perp}{q_\perp},\mu^2\right) \, ,
\label{eq:fragment}
\end{equation}
where $D_{g\rightarrow h}(z,\mu^2)$ is the fragmentation function denoting the probability, at the scale $\mu^2$, that a gluon fragment into 
a hadron carrying a fraction $z$ of its transverse momentum. For simplicity, we only consider here the probability for gluons fragmenting into the 
hadron. The lower limit of the integral can be determined from the kinematic requirement that 
$x_{1,2}\leq 1$--we obtain,
\begin{equation}
z_{\rm min.} = \frac{q_\perp}{m_N}\,e^{y-Y_{\rm beam}} \, .
\label{eq:min-frac}
\end{equation}
if $z_{\rm min.}$ were zero, the effect of including fragmentation effects would simply be to multiply eq.~(\ref{eq:fragment}) by an overall constant 
factor. At lower energies, the typical value of $q_\perp$ is small for a fixed $y-Y_{\rm beam}$; the value of $z_{\rm min.}$ is quite low. However, as 
the center of mass energy is increased, the typical $q_\perp$ value grows slowly with the energy. This has the effect of raising $z_{\rm min.}$ for 
a fixed $y-Y_{\rm beam}$, thereby lowering the value of the multiplicity in eq.(\ref{eq:fragment}) for that $y-Y_{\rm beam}$. Note further that 
eq.~(\ref{eq:min-frac}) suggests that there is a kinematic bound on $q_\perp$ as a function of $y-Y_{\rm beam}$--only very soft gluons can 
contribute to the inclusive multiplicity. 

\begin{figure}[hbt]
\centerline{\epsfig{file=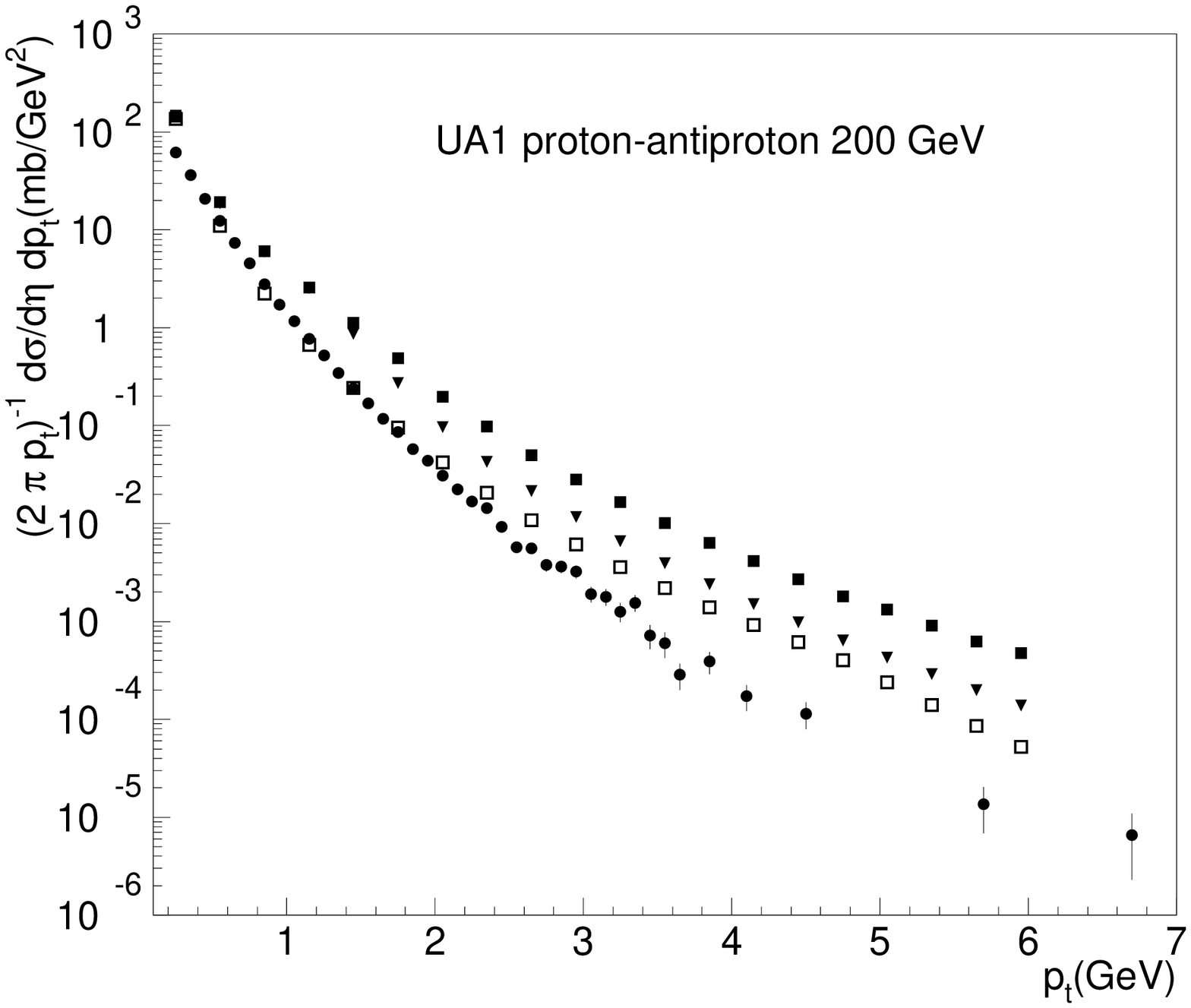,width=0.9\textwidth}}
\caption{ $p_\perp$ distribution from eq.~(\ref{eq:ktfact}) with MV
(full squares) and GBW (full triangles) initial conditions.  The MV
initial condition--with the fragmentation function included--is denoted
by the open squares. The distribution is averaged over the
rapidity region $y=0.0-2.5$, to compare with data (in 200
GeV/nucleon proton-antiproton collisions in the same pseudo-rapidity
range) on charged hadron $p_\perp$ distributions from the UA1
collaboration: full circles.}
\label{fig:pt_pp}
\end{figure}

In figure~\ref{fig:pt_pp} we illustrate $p_\perp$ distributions obtained from the
MV input compare to the UA1 data \cite{UA1}.  We compare the
calculation with and without the fragmentation function.  The
fragmentation function has been taken from \cite{KKPFRAG}.  Clearly
the ``bare" MV model does not describe the data at large $k_\perp$
because it does not include fragmentation function effects which, as discussed, make
the spectrum less flat.  In contrast, because the $k_\perp$ spectrum of
the GBW model dies exponentially at large $k_\perp$, this
``unphysical" $k_\perp$ behavior mimics the effect of fragmentation
functions--see figure \ref{fig:pt_pp}. Hence extrapolations of this
model, as shown in figure~\ref{fig:4} give a more reasonable looking
result. Similar conclusions were reached previously in Ref.~\cite{Szczurek}.

\begin{figure}[hbt]
\centerline{\epsfig{file=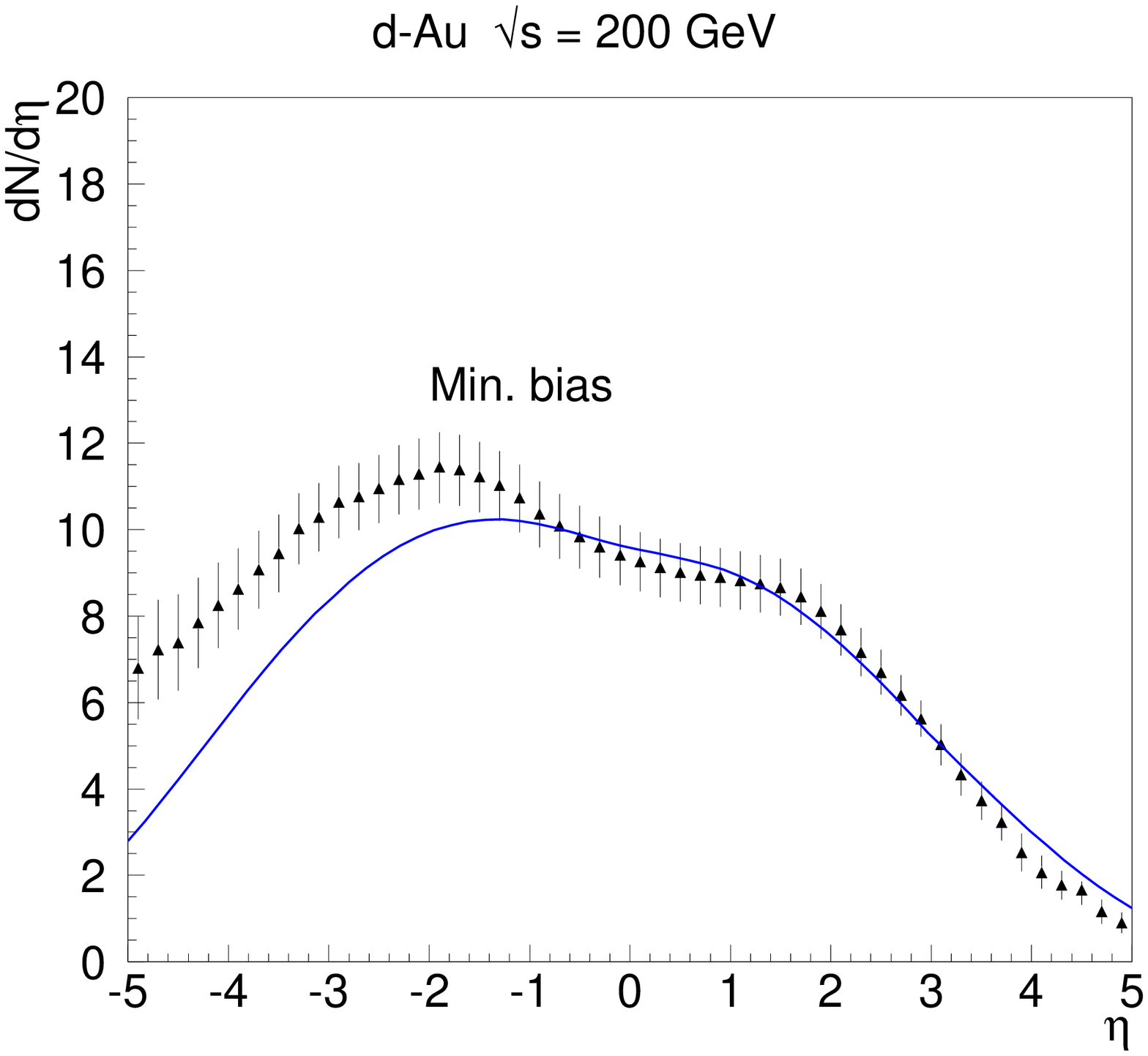,width=0.9\textwidth}}
\caption{
Comparison of computations to minimum bias deuteron-gold data 
at RHIC~\cite{PHOBOS}. Calculation done using the MV model as a starting distribution
for the evolution.}
\label{fig:6}
\end{figure}

We next compute the pseudo-rapidity distribution in 
deuteron-gold collisions.  In figure~\ref{fig:6} we show the result for
the calculation compared with the dA data~\cite{PHOBOS}.  The unintegrated
gluons were extracted from the pp and AA data. The overall shape of
the distribution matches well on the deuteron side with the
minimum-bias data.  The disagreement on the nuclear fragmentation side is easy to understand since, as
mentioned earlier, it requires a better implementation of nuclear geometry effects. Similar conclusions were 
reached in Ref.~\cite{KLN} in their comparisons to the RHIC deuteron-gold data.

\begin{figure}[hbt]
  \centerline{\epsfig{file=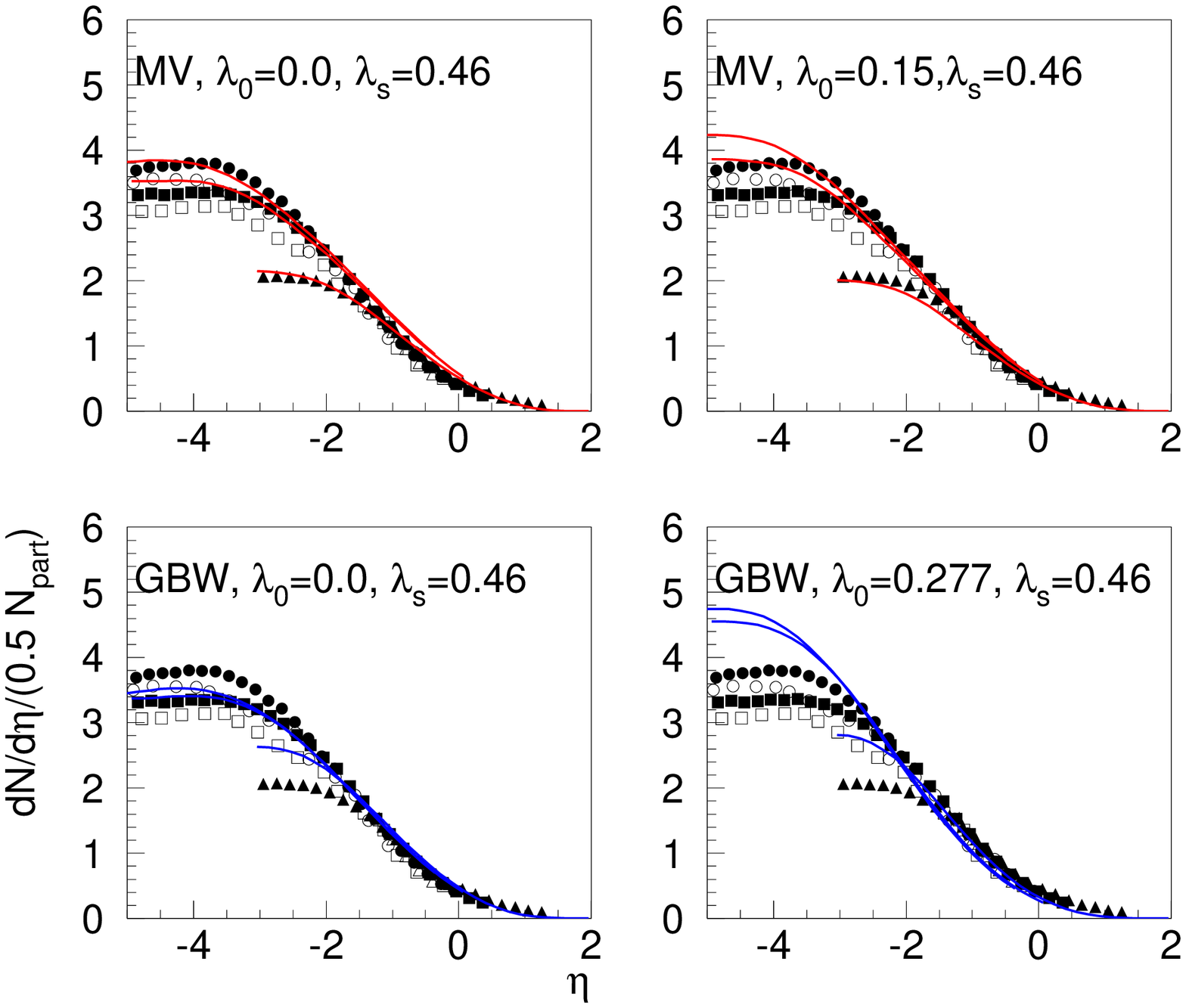,width=0.9\textwidth}}
\caption{ Pseudorapidity distributions normalized by the number of
participants for charged hadrons in gold-gold collisions from the
PHOBOS collaboration at energies $19.6, 130, 200 \, {\rm GeV}$ (filled
triangles , squares and circles), BRAHMS collaboration at energies
$130, 200 \, {\rm GeV}$ (open squares and circles) . The data from the 
STAR collaboration at energy $62.4 \, {\rm GeV}$ (open triangles) are not visible on this plot but can be seen more clearly in fig.~\ref{fig:auauext}. 
Upper plots: initial distributions from the MV model; lower plots:
initial distributions from the GBW model.  }
\label{fig:5}
\end{figure}

We now apply our considerations to central nucleus--nucleus collisions.  In
figure~\ref{fig:5} we present fits to data on the pseudo-rapidity
distributions in gold--gold collisions from the PHOBOS, BRAHMS and STAR
collaborations.  The data~\cite{PHOBOS} are for $\sqrt{s} =19.6,130,
200 \, {\rm GeV}$ and the BRAHMS data~\cite{BRAHMS} are for $\sqrt{s}=130,200 \,
{\rm GeV}$.  A reasonable description of limiting fragmentation is
achieved in this case as well. One again has discrepancies in the
central rapidity region as in the pp case. As mentioned previously, a
natural explanation for this discrepancy is the violation of
$k_\perp$-factorization. It is expected that these violations decrease
the multiplicity in this regime~\cite{KNV,Balitsky2} relative to
extrapolations using $k_\perp$-factorization. We find that values of
$Q_s^A\approx 1.3$ GeV for the saturation scale give the best fits.
This value is consistent with other
estimates~\cite{KNV,KL,KowalskiTeaney}.  Apparently the gold-gold data
are better described by the calculations which have
$\lambda_0=0.0$. This might be related to the difference in the large
$x$ distributions in the proton and nucleus.  Also, slightly higher
values of $\lambda_s$ are preferred.  This variation of parameters
from AA to pp case might be also connected with the fact that in our
approach the impact parameter is integrated out,  so that there is no detailed 
information on the nuclear geometry.  A more detailed calculation with the 
impact parameter dependence taken into account is left for future investigations.

\begin{figure}[hbt]
  \centerline{\epsfig{file=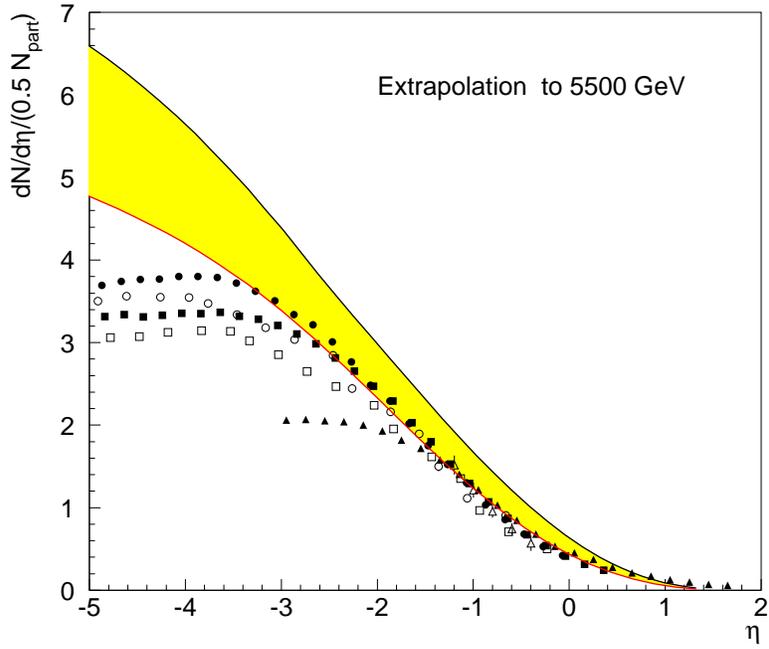,width=0.9\textwidth}}
  \caption{ Extrapolation of calculations for gold-gold collisions
    shown in fig.~\ref{fig:5} to the LHC energy $\sqrt{s}=5500 \; {\rm
      GeV}$/ nucleon. For comparison, the same data at lower energies
    are shown. (See fig.~\ref{fig:5}.) The band is an estimate of the
    systematic uncertainty of our approach. Its lower border
    corresponds to the GBW input with $\lambda_0=0,\lambda_s=0.46$,
    and its upper border to the MV input with
    $\lambda_0=0,\lambda_s=0.46$. }
\label{fig:auauext}
\end{figure}

In fig.~\ref{fig:auauext} we show the extrapolation of two
calculations to higher energy $\sqrt{s}=5500 \; {\rm GeV}$.  We note
that the calculation within the MV model gives results which would
violate the scaling in the limiting fragmentation region by
approximately $25\%$ at larger $y-Y_{\rm beam}$.  This violation is
partly because of the effect of fragmentation functions discussed
previously and in part due to the fact that the integrated parton
distributions from the MV model do not obey Bjorken scaling at large
values of $x$.  In the latter case, the violations are proportional to
$\ln(Q_s^2(x_2))$ as discussed previously. The effects of the former
are simulated by the GBW model-the extrapolation of which, to higher
energies, is shown by the dashed line. The band separating the two
therefore suggests the systematic uncertainity in such an
extrapolation coming from a) the choice of initial conditions and b)
the effects of fragmentation functions which are also uncertain at
lower transverse momenta.
 
\begin{figure}[hbt]
\centerline{\epsfig{file=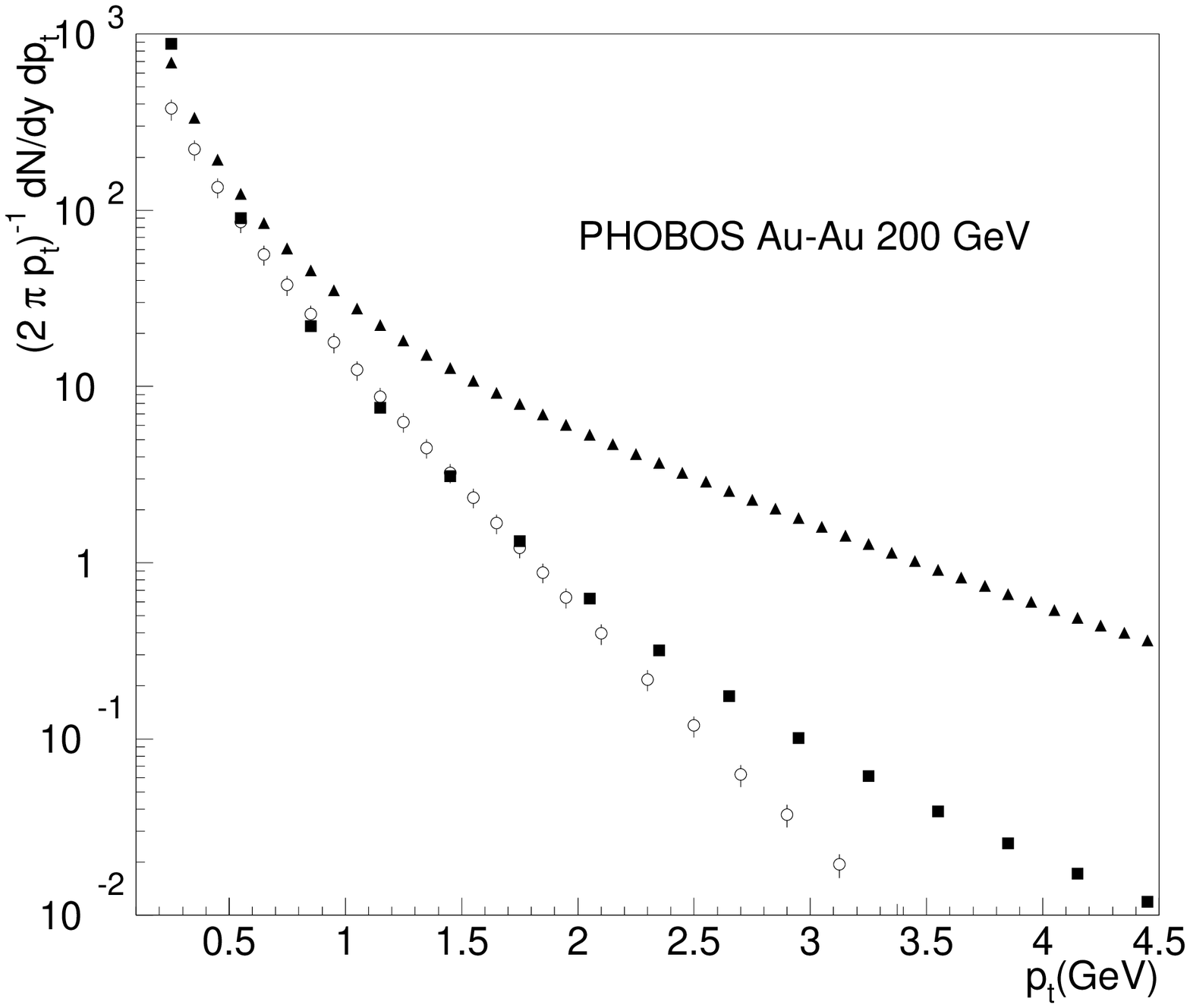,width=0.9\textwidth}}
\caption{ $p_\perp$ distribution from eq.~(\ref{eq:ktfact}) with MV
(full triangles)  initial conditions.  The MV
initial condition with the fragmentation function included is denoted
by the full squares. The distribution is averaged over the
pseudo-rapidity region $y=0.2-1.4$, to compare with data in 200
GeV/nucleon gold-gold collisions (open circles).   }
\label{fig:pt_auau}
\end{figure}

Finally, the transverse momentum distribution in the gold-gold
collisions is presented in fig.~\ref{fig:pt_auau}. The data at
$\sqrt{s}=200 {\rm GeV}$ are from the PHOBOS collaboration
\cite{PHOBOS_pt}. Again, in this case, the calculation without
fragmentation function tends to overshoot the data significantly at
high values of $p_t$. Including fragmentation function
effects~\cite{KKPFRAG} results in a better agreement with the shape of
the $p_t$ distribution as expected. A more detailed study to extract
parameters that give the best chi-squared values will be performed at
a later date.

\section{Conclusions}

We studied here the phenomenon of limiting fragmentation in the Color
Glass Condensate framework. In the dilute-dense (projectile-target)
kinematics of the fragmentation regions, one can derive (in this
framework) an expression for inclusive gluon distributions which is
$k_\perp$ factorizable into the product of ``unintegrated" gluon
distributions in the projectile and target.  From the general formula for 
gluon production(eq.~(\ref{eq:ktfact})), 
limiting fragmentation is a consequence of two factors:
\begin{itemize}
\item Unitarity of the $U$ matrices which appear in the definition of
the unintegrated gluon distribution in eq.~(\ref{eq:unintBGV}).
\item Bjorken scaling at large $x_1$, namely, the fact that the
integrated gluon distribution at large $x$, depends only on $x_1$ and
not on the scale $Q_s(x_2)$. (The residual scale dependence consequently leads to the
dependence on the total center-of-mass energy.)
\end{itemize}
  
Deviations from the limiting curve at experimentally accessible
energies are very interesting because they can potentially teach us
about how parton distributions evolve at high energies. In the CGC
framework, the Balitsky--Kovchegov equation determines the evolution
of the unintegrated parton distributions with energy from an initial
scale in $x$ chosen to be $x_0=0.01$. This choice of scale is inspired
by model comparisons to the HERA data.

We compared our results to data on limiting fragmentation from pp
collisions at various experimental facilities over a wide range of
collider energies, and to collider data from RHIC for deuteron-gold
and gold-gold collisions. We obtained results for two different models
of initial conditions at $x\geq x_0$; the McLerran-Venugopalan model
(MV) and the Golec-Biernat--Wusthoff (GBW) model.  In addition to the
two parameters in the initial conditions ($\lambda_s$ and
$\lambda_0$), we also studied the sensitivity of our results to an
infrared momentum cut-off $m$ (chosen to be the same value in the
$y\rightarrow \eta$ conversion for hadrons).

We found reasonable agreement for this wide range of collider data for
the limited set of parameters. Clearly these can be fine tuned by
introducing further details about nuclear geometry. That would
introduce further parameters but on the other hand there is more data
for different centrality cuts as well-we leave these detailed
comparisons for future studies. In addition, an important effect,
which improves agreement with data, is to account for the
fragmentation of gluons in hadrons. In particular, the MV model, which
has the right leading order large $k_\perp$ behavior, but no
fragmentation effects, is much harder than the data. The latter falls
as a much higher power of $k_\perp$. Since even rapidity distributions
at higher energies data are more sensitive to larger $k_\perp$, we
expect this discrepancy to show up in our studies of limiting
fragmentation and indeed it does. We noticed that taking this into
account lead to much more plausible extrapolations of fits of existing
data to LHC energies. This ``gluon fragmentation" contribution also
suggests that the ``flat" deviations that we found (for fits with MV
initial conditions) for $\lambda_s = 0.46$ ($\lambda_{\rm BK}=0.28$)
are diminished.

Clearly, the procedure employed in this paper needs further improvements.
One of them is the impact parameter dependence of the unintegrated
gluon distribution functions.  We discussed briefly ``gluon
fragmentation" effects (with different fragmentation function sets)
which need to be taken into account. Furthermore, large $x$
distributions also need to be better constrained and consistency with
computations of other final states in the CGC framework established.
Finally, the factorization formula used in this paper involves only
gluons;  for  more realistic estimates, one should also include quark
distributions in this framework.

\section*{Acknowledgments}
This work was inspired by discussions with A. Bia\l{}as and
L. McLerran and by a seminar at Saclay by A. Bia\l{}as that was
attended by FG and RV. We would also like to thank M. Baker and
R. Noucier of the PHOBOS collaboration and M. Murray and F. Videbaek
of the BRAHMS collaboration for their help with their respective data
sets and for useful comments. We thank B. Mohanty for bringing the STAR data 
on limiting fragmentation to our attention. Finally, FG wishes to thank the hospitality of
the Nuclear Theory group at BNL. AMS and RV were supported by DOE
Contract No. DE-AC02-98CH10886. AMS was also supported by the Polish
Committee for Scientific Research, KBN grant No. 1 P03B 028 28.


\end{document}